\documentclass[apjl]{emulateapj}
\usepackage{subfigure}
\usepackage{wrapfig}

\def\lsim{\mathrel{\rlap{\lower4pt\hbox{\hskip1pt$\sim$}}
    \raise1pt\hbox{$<$}}}                
\def\gsim{\mathrel{\rlap{\lower4pt\hbox{\hskip1pt$\sim$}}
    \raise1pt\hbox{$>$}}}                

\shorttitle{High-efficiency Autonomous Laser Adaptive Optics}
\shortauthors{Baranec et al.}
\begin{document}
\title{High-efficiency autonomous laser adaptive optics}

\author{Christoph Baranec\altaffilmark{1}, Reed Riddle\altaffilmark{2}, Nicholas M. Law\altaffilmark{3}, A. N. Ramaprakash\altaffilmark{4}, Shriharsh Tendulkar\altaffilmark{2}, Kristina Hogstrom\altaffilmark{2}, Khanh Bui\altaffilmark{2}, Mahesh Burse\altaffilmark{4}, Pravin Chordia\altaffilmark{4}, Hillol Das\altaffilmark{4}, Richard Dekany\altaffilmark{2}, Shrinivas Kulkarni\altaffilmark{2}, and Sujit Punnadi\altaffilmark{4}}

\altaffiltext{1}{Institute for Astronomy, University of Hawai`i at M\={a}noa, Hilo, HI 96720-2700, USA; baranec@hawaii.edu}
\altaffiltext{2}{Division of Physics, Mathematics, and Astronomy, California Institute of Technology, Pasadena, CA, 91125, USA}
\altaffiltext{3}{Department of Physics and Astronomy, University of North Carolina at Chapel Hill, Chapel Hill, NC 27599-3255, USA}
\altaffiltext{4}{Inter-University Centre for Astronomy \& Astrophysics, Ganeshkhind, Pune, 411007, India}

\begin{abstract}
As new large-scale astronomical surveys greatly increase the number of objects targeted and discoveries made, the requirement for efficient follow-up observations is crucial. Adaptive optics imaging, which compensates for the image-blurring effects of Earth's turbulent atmosphere, is essential for these surveys, but the scarcity, complexity and high demand of current systems limits their availability for following up large numbers of targets. To address this need, we have engineered and implemented Robo-AO, a fully autonomous laser adaptive optics and imaging system that routinely images over 200 objects per night with an acuity 10 times sharper at visible wavelengths than typically possible from the ground. By greatly improving the angular resolution, sensitivity, and efficiency of 1--3 m class telescopes, we have eliminated a major obstacle in the follow-up of the discoveries from current and future large astronomical surveys. 
\end{abstract}


\maketitle

\def \Kepler {\textit{Kepler}}
\section{Introduction}

Extremely large astronomical surveys necessitate highly efficient follow-up instrumentation to fully characterize the large numbers of discoveries made \citep{K12}. Adaptive optics imaging, which compensates for the image-blurring effects of EarthÕs turbulent atmosphere, is essential for these surveys: e.g., validating exoplanet candidates \citep{Morton11, M14, Law14}, detecting and probing the properties of unresolved binary star systems \citep{L7A, Metchev09, Bowler12, Terziev13}, and spatially locating and measuring supernovae in their host environments \citep{Ofek07, Li11, Cao13}. Current laser adaptive optics systems offer the most flexible high-resolution-imaging capability, but their scarcity and complexity along with the high demand for observing time on large apertures limits their suitability for following up large numbers of targets \citep{Hart10, Davies12}. To address this need, we have engineered and implemented Robo-AO, a fully autonomous laser adaptive optics and imaging system that routinely images over 200 objects per night with an acuity 10 times sharper at visible wavelengths than typically possible from the ground. We have used Robo-AO to complete an adaptive optics survey of 715 \Kepler ~exoplanet host candidates, revealing tentative evidence that short-period giant exoplanets are more likely than other exoplanets to be found in wide stellar binary systems \citep{Law14}. We have also used Robo-AO observations to validate individual \Kepler ~planet candidates \citep{Swift13, Muirhead14}, to investigate exotic binary systems \citep{Law12, Muirhead13}, and to search for companions to thousands of nearby stars. 

Robo-AO systems offer the opportunity to upgrade the more than one hundred 1-3 m class telescopes that are currently operating without the benefits of adaptive optics. By greatly improving the angular resolution, sensitivity, and efficiency of 1--3 m class telescopes, we have eliminated a major obstacle in the follow-up of current and future large astronomical surveys. In this Letter we describe the Robo-AO instrument (Section~\ref{sec:methods}), describe the initial results from the system and its delivered image quality (Section~\ref{sec:iq}), and detail the future plans for the system (Section~\ref{sec:plans}).

\section{Methods}
\label{sec:methods}

The Robo-AO instrument, mounted to the robotic 1.5-m telescope at Palomar Observatory \citep{Cenko06}, comprises several main systems: a laser-launch system; a set of support electronics; a Cassegrain instrument package that houses a high-speed optical shutter, wavefront sensor, wavefront corrector, science instrument and calibration sources; and a single computer that controls the entire system (\citealt{Baranec2013}; summarized in Table~\ref{tab:survey_specs}). 

\begin{deluxetable}{ll}

\tablecaption{\label{tab:survey_specs}Palomar Robo-AO Specifications}
\label{tab:specs}
\tabletypesize{\footnotesize}

\startdata
\hline
\hline
\noalign{\vskip 1mm}   
Telescope     & Palomar 1.5-m telescope \\
Science camera & Andor iXon DU-888 \\
EMCCD detector & E2V CCD201-20 \\
Read-noise (without EM gain) & 47 $e^{-}$ \\
EM gain, selectable & 300, 200, 100, 50, 25 \\
Effective read-noise & 0.16, 0.24, 0.48, 0.96, 1.9 $e^{-}$ \\
Full-frame-transfer readout & 8.6 frames per second \\
Detector format & 1024$^2$ 13 $\mu$m pixels\\
Field of view & 44\arcsec $\times$ 44\arcsec\\
Pixel scale & 43.1 milli-arcsec per pixel\\
Observing filters & Sloan $g'$, $r'$, $i'$, $z'$, and $>$600 nm \\
Tip-tilt guide star range &	2 $<$ $m_V$ $<$ 16 \\
Residual RMS wavefront error & 141-218 nm \\
Strehl ratio at $i'$-band & $26\%-4\%$ \\
LGS-AO overheads per target & 40 $-$ 42 s (typical) \\
Average telescope slew time & 40 s \\

\hline

\end{deluxetable}

\begin{figure}[b]
  \centering
  \resizebox{1.0\columnwidth}{!}
   {
    \includegraphics{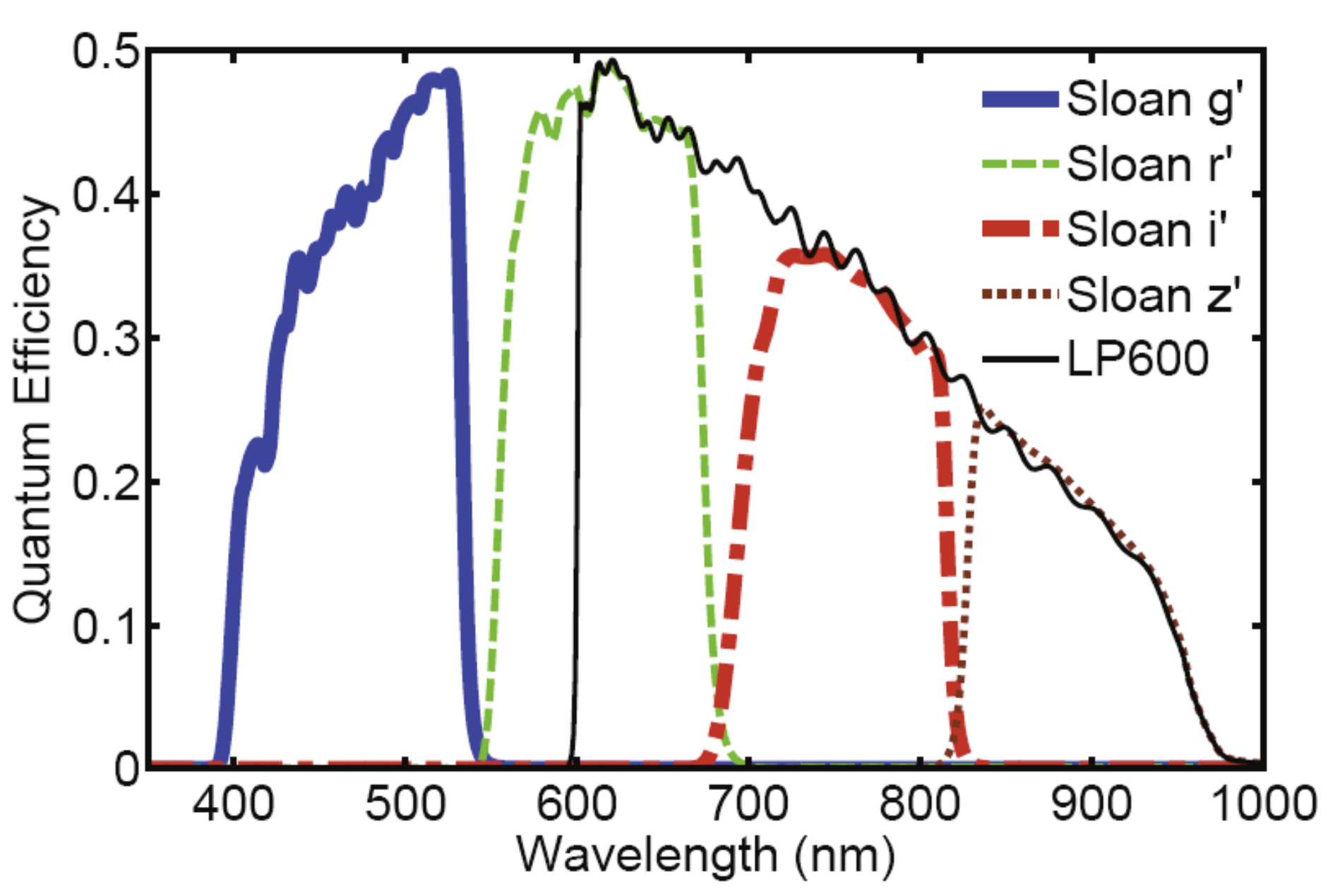}
   }
   \caption{Quantum efficiency of Robo-AO as a function of wavelength in different observing filters. The curves are generated from measured reflection and transmission data from all optical components with the exception of the primary and secondary of the 1.5-m telescope which are assumed here to be ideal bare aluminum.\label{fig:one}}

\end{figure}

\begin{figure*}
  \centering
  \resizebox{0.95\textwidth}{!}
   {
    \includegraphics{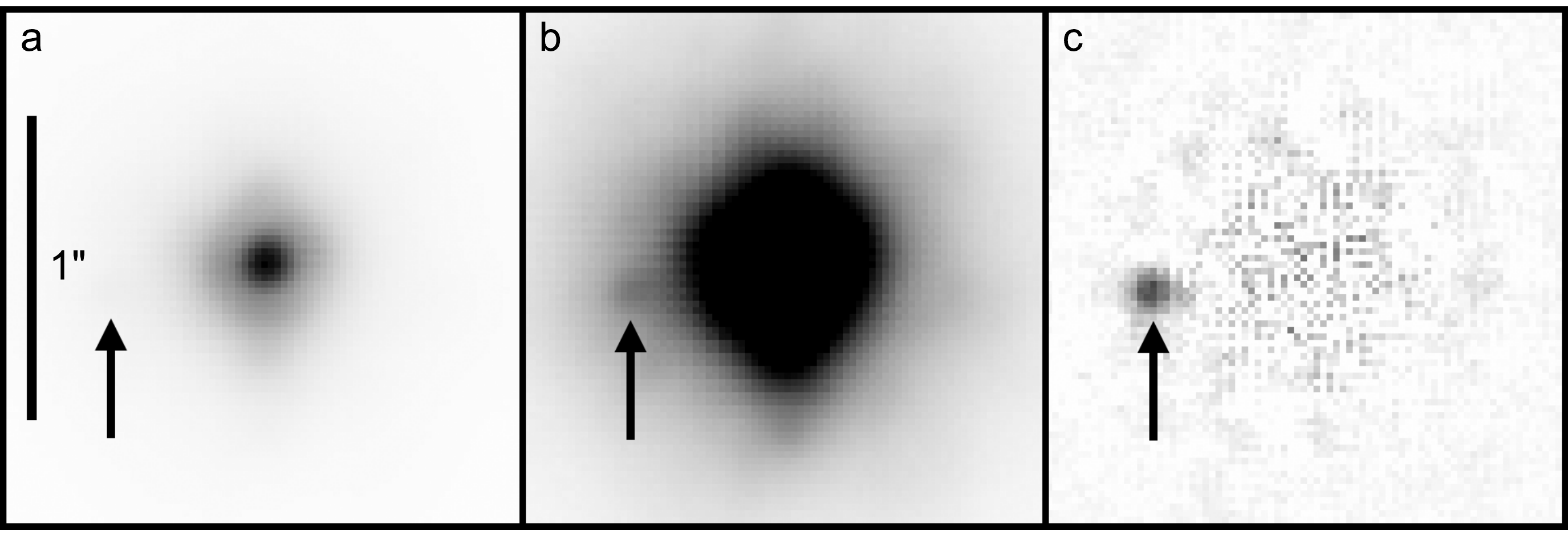}
   }
   \caption{Robo-AO visible-light observation of a star revealing a faint close companion as indicated by an arrow. (a), linear scaling to the peak intensity of the stellar PSF. (b), linear scaling to 10\% of the peak intensity of the stellar PSF. Quasi-static instrumental and atmospheric speckles are revealed in the stellar halo. (c), image after PSF subtraction, linear scaling to 2\% of the peak intensity of the primary stellar PSF. Speckles are suppressed and a faint companion is revealed.\label{fig:two}}

\end{figure*}

\begin{figure*}
  \centering
  \resizebox{1.0\textwidth}{!}
   {
    \includegraphics{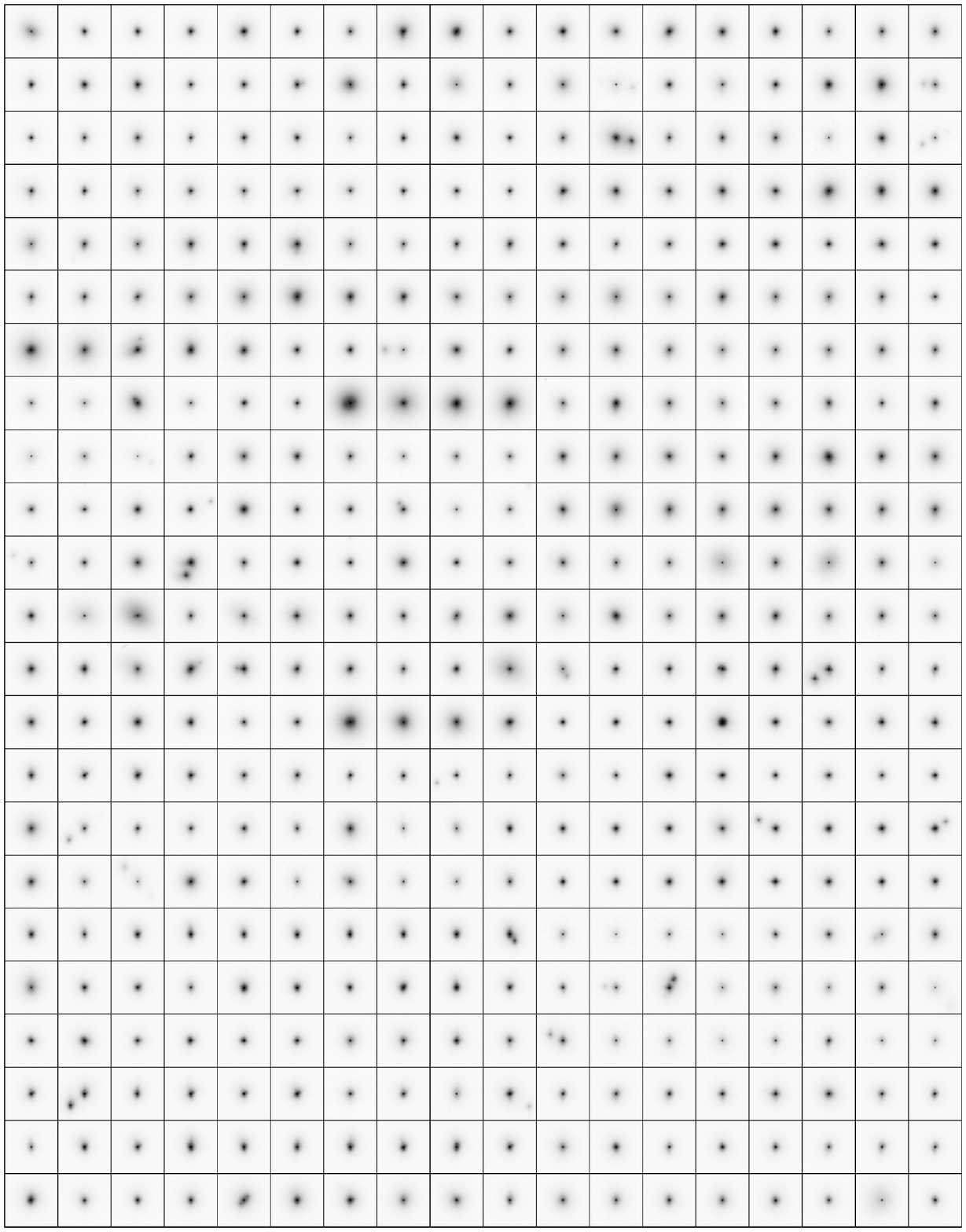}
   }
   \caption{Robo-AO adaptive optics images of 414 stars within 25 parsecs of the Sun. Each square represents a 3\arcsec $\times$ 3\arcsec area and 90 s of integration in $i'$-band, $\overline{\lambda}$=765 nm.\label{fig:three}}
   \label{fig:binaries_summary}
\end{figure*}

The laser-launch system comprises a pulsed, 12 W, $\lambda$ = 355 nm laser beam, coaligned with the bore-sight of the principal telescope, and focuses a seeing-limited beam waist with a 15-cm projection aperture to a line-of-sight distance of 10 km. As with other ultraviolet laser systems propagating into navigable airspace (e.g., \citealt{Thompson02, Tokovinin12}), control measures to avoid illuminating aircraft are not required because the laser beam is unable to flash-blind pilots or produce biologically hazardous radiation levels during momentary exposures. The instrument package located at the Cassegrain telescope focus uses an opto-electrically shuttered Shack--Hartmann wavefront sensor to record the light returning to the telescope from the laser focus over a backscatter range of 450 m. The signal-to-noise ratio of each Shack--Hartmann slope measurement ranges from 6 to 10 depending on Zenith angle and seeing conditions (6.5 electrons of detector noise in each of four binned pixels with a signal ranging from 100 to 200 photoelectrons per subaperture). The measured tip and tilt signals are used to stabilize the laser pointing with an uplink beam-steering mirror, while the remaining atmospheric wavefront aberrations, excluding stellar image displacement, are compensated with a 12 $\times$ 12 actuator, 3.5 $\mu$m stroke, micro-electromechanical deformable mirror. The adaptive optics control loop operates at the 1.2 kHz frame rate of the Shack--Hartmann detector, with an effective wavefront control bandwidth of 90--100 Hz.

Robo-AO is the first astronomical adaptive optics system to operate autonomously; a master robotic sequencer controls the telescope, adaptive optics system, laser, filter wheel, and science camera; executing all operations that otherwise would have been performed manually, allowing greatly improved observing efficiency. The software to control each hardware subsystem was developed as a set of individual modules in C++, and small standalone test programs have been created to test each of the hardware interfaces. This modular design allows the individual subsystems to be stacked together into larger modules, which can then be managed by the master robotic sequencer.

The execution of an observation starts with a query to an intelligent queue scheduling program that selects a target. The robotic sequencer will then point the telescope, while simultaneously selecting the appropriate optical filter and configuring the science camera, laser and adaptive optics system. A laser acquisition process to compensate for differential pointing between the telescope and laser projector optical axes, caused by changing gravity vectors, begins once the telescope has completed pointing at the new target. A search algorithm acquires the laser by moving the uplink steering mirror in an outward spiral pattern from center until 80\% of the wavefront sensor subapertures have met a flux threshold of 75\% of the typical laser return flux. Simultaneous with the laser acquisition process, the science camera is read out for 20 s with no adaptive optics compensation and with the deformable mirror fixed to obtain a contemporaneous estimate of the seeing conditions through the telescope. Upon completion of laser acquisition, a new wavefront sensor background image is taken, the adaptive optics correction is started and an observation with the science camera begins.

During an observation, telemetry from the adaptive optics loop is used to maintain telescope focus and detect significant drops in laser return flux. Slow drifts in the focus mode of the deformable mirror are measured and offloaded to the secondary mirror to preserve the dynamic range of the deformable mirror. Focus on the deformable mirror is measured by projecting the commanded actuator values to a model Zernike focus mode. A median of the last 30 focus values, measured at 1-s intervals, is calculated. If the magnitude of this value exceeds 220 nm peak-to-valley surface of focus on the deformable mirror, equivalent to a displacement of 20 ?m of the Palomar 1.5-m telescope secondary mirror, then the secondary is commanded to change focus to null out this value. Focus corrections may not be applied more than once every 30 s and are restricted to less than 50 $\mu$m of total secondary motion to avoid runaway focus. The laser return flux is also measured at simultaneous 1-s intervals. If the laser return drops below 50 photoelectrons per subaperture on the wavefront sensor for more than 10\% of the values used to calculate the median focus, e.g., due to low-altitude clouds or extremely poor seeing (greater than 2\farcs5), any focus correction is ignored due to the low certainty of the measurement. Additionally, if the return stays below 50 photoelectrons per subaperture for five consecutive seconds, the observation is immediately aborted, the target is marked as ``attempted but not observed'' in the queue, and a new target is selected for observation.

The intelligent queue is able to pick from all targets in a directory structure organized by scientific program, with observation parameters defined within Extensible Markup Language files. The queue uses an optimization routine based on scientific priority, slew time, telescope limits, prior observing attempts, and laser-satellite avoidance windows to determine the next target to observe. In coordination with US Strategic Command (USSC), we have implemented measures to avoid laser illumination of satellites. To facilitate rapid follow-up observations we have developed new de-confliction procedures which utilize the existing USSC protocols to open the majority of the overhead sky for possible observation without requiring preplanning. By requesting predictive avoidance authorization for individual fixed azimuth and elevation ranges, as opposed to individual sidereal targets, Robo-AO has the unique capability to undertake laser observations of the majority of overhead targets at any given time.

The 1.5-m telescope takes, on average, 40 s to point to a new target, and Robo-AO requires 40 to 42 s for the laser acquisition before starting a new observation; this is a substantial improvement over manually operated laser adaptive optics systems that typically require 5 to 35 minutes \citep{Amico10, Minowa12, Neichel12, Wizinowich13}. This high cadence leads to the routine nightly observation of over 200 targets.

Science observations are made with an electron multiplying CCD camera with a 44\arcsec square field of view and 0\farcs043 pixel scale. Two six-position filter wheels are directly before the camera and are equipped with Sloan $g'$-, $r'$-, $i'$- and $z'$-band filters \citep{York2000} as well as a long-pass filter cutting on at 600nm (LP600 hereafter), a blocking filter and room for four additional filters. The quantum efficiency of the entire system as a function of wavelength in the different filters is presented in Figure~\ref{fig:one}. A selectable electron multiplication gain factor of 25, 50, 100, 200 or 300 can be enabled to reduce the non-amplified read noise of 47.8 $e^{-}$ to as low as 0.16 $e^{-}$ at the cost of dynamic range. The camera is read out continually at a frame rate of 8.6 Hz during science observations, allowing image displacement, that cannot be measured using the laser system \citep{Rigaut92}, to be removed in software based on the position of a $m_V \leq 16$ guide star within the field of view.

Upon completion of an observation, the data is compressed and archived to a separate computer system where the data is immediately processed. A data reduction pipeline \citep{Law09b, Law14} corrects each of the recorded frames for detector bias and flat-fielding effects, and automatically measures the location of the guide star in each frame. The region around the star is up-sampled by a factor of four using a cubic interpolation, and the resulting image is cross-correlated with a diffraction-limited point spread function for that wavelength. This has been shown to obtain much higher quality results than centroid or brightest-pixel alignment (e.g., \citealt{Law06}). The frame is then shifted to align the position of greatest correlation to that of the other frames in the observation, and the stack of frames is coadded using the Drizzle algorithm \citep{Fruchter02} to produce a final high-resolution output image sampled at twice the resolution of the input images. The rate at which final images are processed lags only slightly behind the data capture rate: a full nightÕs set of data is typically finished before the next night of observing.

\section{Results}
\label{sec:iq}

For science programs which require the detection and contrast ratio measurement of closely separated objects, an additional point-spread-function (PSF) subtraction and analysis pipeline can be started upon completion of the data reduction pipeline. This pipeline distinguishes astrophysical objects from residual atmospheric and instrumental wavefront errors and corresponding speckles in the image plane using a modified Locally Optimized Combination of Images \citep{L07b} algorithm. The algorithm selects a combination of similar PSFs from the hundreds of other observations of similar targets during that night. The combination of PSFs is used to create a model PSF which is then subtracted from each image and potential companions are flagged. Figure~\ref{fig:two} shows an example of the PSF subtraction of an automatically reduced observation that reveals a source at 0\farcs53 away at a contrast ratio of $1/45$ with respect to the primary star. The PSF subtraction pipeline achieves photon-noise limited contrast of $1/100$ at angular distances greater than 0\farcs2 from typical target stars. While this process can be fully automated, we have thus far adopted a strategy to manually verify intermediate and final results of the analysis pipeline process to ensure accurate results, e.g., initially confirming the guide star position and visually checking detected companions.

\subsection{Science Observations}

Robo-AO has already performed a wide range of surveys, including a comprehensive adaptive-optics survey for stellar binaries in the solar neighborhood. We have used Robo-AO to perform follow-up observations of over 3000 stars that have been identified by the Research Consortium on Nearby Stars survey (RECONS; e.g., \citealt{Henry94}) to be within 25 parsecs of our Sun. Our observations will augment ongoing nearby stellar multiplicity studies by discovering stellar companions in the critical range of 1--100 AU, which are otherwise too close to be detected by seeing limited observations, or too long in orbital period to be detected by radial velocity or astrometric methods. Figure~\ref{fig:three} shows Robo-AO observations of the first 414 of these nearby stars completed in just 2.5 nights. In this sample alone, we have discovered 37 close binary systems, and identified a possible new nearby triple star system; observations at later epochs will confirm these associations. As of this writing, Robo-AO has completed 490 hr of fully robotic operations during 88 nights of allocated telescope time, of which 267 hr were open-shutter science observing time. In total, the system has completed approximately 10,000 science observations, with typical exposure times ranging from 30 s to 3 minutes each.

\begin{figure}
  \centering
  \resizebox{1.0\columnwidth}{!}
   {
    \includegraphics{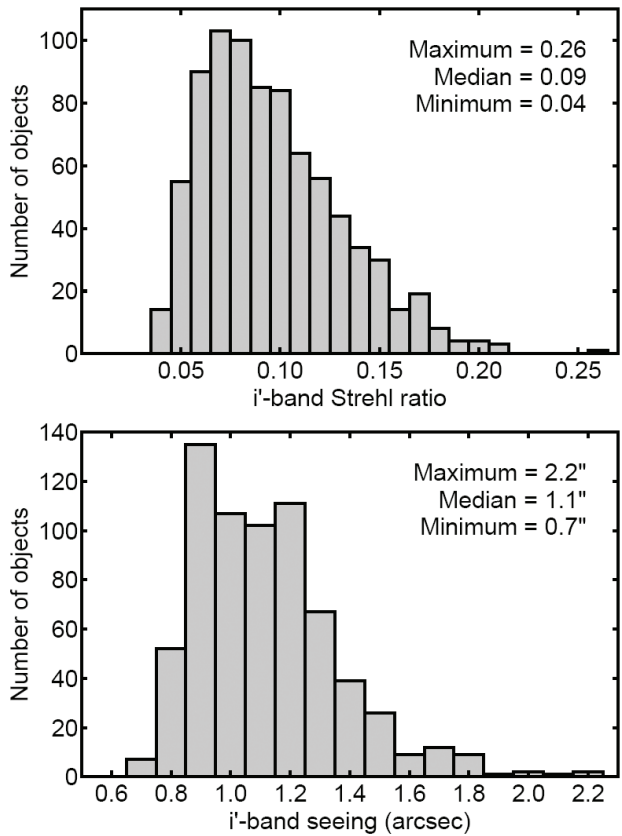}
   }
   \caption{Summary of adaptive optics corrected $iÕ$-band Strehl ratios (left) and contemporaneous seeing measurements (right).\label{fig:four}}

\end{figure}

\subsection{Image Quality}

The sharpest achieved image widths over the course of initial commissioning were measured to be 0\farcs25, 0\farcs10, 0\farcs12 and 0\farcs14 in the Sloan $g'$-, $r'$-, $i'$- and $z'$-bands respectively; in $r'$, $i'$ and $z'$ these resolutions correspond to the diffraction limit of the 1.5-m telescope. The long-term $i'$-band image quality improvement of the adaptive optics correction was measured as part of our program to search for stellar binaries in the solar neighborhood and is summarized in Figure~\ref{fig:four}. The 812 adaptive optics corrected observations reported here, each 90 or 120 s in total exposure time, are drawn from a total of 61 different nights, spanning 2012 June 17 to 2013 October 23. They exclude observations from other science programs, identified multiple stars, objects too faint for the automatic image registration algorithm (see \citealt{Law12} for an example of observations of stars $m > 16$), or objects that are too bright, typically $m_V < 2$. For comparison to observing conditions, seeing limited data were taken during the time of laser acquisition prior to each of the adaptive optics corrected observations after 2012 July, comprising 682 contemporaneous seeing measurements. While our median $i'$-band seeing value of 1\farcs1 may underestimate the true seeing due to the short 20-s exposure times used, it is comparable to the average $\sim$1\farcs1 $R$-band seeing found by \cite{Cenko06}. Under the measured observing conditions, we find the residual root-mean-square optical wavefront errors to be in the range of 141 to 218 nm, with a median of 189 nm, equivalent to $i'$-band Strehl ratios between 26\% and 4\%, with a median of 9\%. Significant image sharpening is achieved under the vast majority of conditions, and only as the seeing exceeds 1\farcs6 does the delivered image quality fall below 5\% Strehl.

\section{Future Plans}
\label{sec:plans}

New low-noise infrared cameras will soon be tested with the Palomar Robo-AO instrument which, once deployed, will enhance the effective near-infrared sensitivity of a 1.5-m telescope to that of a seeing limited 4-m telescope, and enable deeper visible-light imaging using adaptive-optics sharpened infrared tip-tilt guide sources. Additional instrumentation, e.g., an integral field spectrograph to quickly classify supernovae in crowded fields, can also be mounted to external instrument ports to take advantage of the automated adaptive optics correction.

Future Robo-AO systems on dedicated telescopes will be able to complete tens of thousands of high acuity science observations in a single year, making statistically meaningful discoveries during the follow-up characterization of large surveys, and will be able to respond immediately to observation requests from transient discovery machines. We have begun work on two new facility Robo-AO systems, which we expect to be operational in 2016-2017, one for the IUCAA Girawali Observatory 2-m telescope in Maharashtra, India, and another for the University of Hawai`i 2.2-m telescope on Maunakea in Hawai`i. We intend to use the new systems to image transiting exoplanet host candidates discovered by the \Kepler ~K2 mission \citep{Howell14} and the Transiting Exoplanet Satellite Survey \citep{Ricker14}, along with candidates produced by ground-based surveys (e.g., \citealt{Law13S}). By identifying and measuring the components contributing to the photometric light curves we will be able to refine the transit properties and validate thousands of potential exoplanet candidates en masse. We will also develop an interruptible queue and partner with transient surveys, e.g., the Asteroid Terrestrial-impact Last Alert System \citep{Tonry11} and the Palomar Transient Factory \citep{Law09a}, to minimize the time between discovery and characterization of rapidly changing transient events.

\acknowledgments

We thank the anonymous referee for their particularly useful suggestions. We thank the staff of Palomar Observatory for their support in the deployment of the Robo-AO system on the 1.5-m telescope. The Robo-AO system is supported by collaborating partner institutions, the California Institute of Technology and the Inter-University Centre for Astronomy and Astrophysics, by the National Science Foundation under Grant Nos. AST-0906060, AST-0960343, and AST-1207891, by a grant from the Mt. Cuba Astronomical Foundation and by a gift from Samuel Oschin. C.B. acknowledges support from the Alfred P. Sloan Foundation. C.B., R.R. and N.M.L. wrote the paper. C.B. led the project. R.R. led the robotic software development. N.M.L. acted as project scientist and developed the data reduction pipeline. All authors contributed to the development of the Robo-AO instrument.

{\it Facility:} \facility{PO:1.5m (Robo-AO)}

\bibliographystyle{apj.bst}
\bibliography{references.bib}

\end{document}